\documentclass[pra,showpacs,twocolumn]{revtex4}
\usepackage{amsmath}
\usepackage{epsfig}
\usepackage{dcolumn}
\usepackage{bm}

\newcommand{\ket}[1]{\left\vert#1\right\rangle}

\newcommand{\exv}[1]{\left\langle#1\right\rangle}

\begin{document}
\title{Quantum Computing in the Presence of Spontaneous Emission
By a Combined Dynamical Decoupling and Quantum Error Correction Strategy}
\author{K. Khodjasteh}
\affiliation{Department of Physics,
  University of Toronto, 60 St.George Street Toronto, ON M5S 1A7,
  Canada.}
\author{D.A. Lidar}
\affiliation{Chemical Physics Theory
  Group, University of Toronto, 80 St. George St., Toronto, ON M5S
  3H6, Canada.}

\begin{abstract}
A new method for quantum computation in the presence of spontaneous
emission is proposed. The method combines strong and fast (dynamical
decoupling) pulses and a quantum error correcting code that encodes
$n$ logical qubits into only $n+1$ physical qubits. Universal,
fault-tolerant, quantum computation is shown to be possible in this
scheme using Hamiltonians relevant to a range of promising proposals
for the physical implementation of quantum computers.
\end{abstract}
\maketitle

\section{Introduction}
Decoherence \cite{Giulini:book} remains the most daunting obstacle to
the realization of quantum information processing, coherent control,
and other applications requiring a high degree of quantum
coherence. As quantum computation (QC) moves into the experimental
realm it becomes increasingly important to design methods for
overcoming this main obstacle to realization, that are tailored to
particular systems and the resulting errors that afflict them.  Here
we show how to perform universal, fault-tolerant QC in the presence of
decoherence due to \emph{spontaneous emission} (SE). Since SE is a
consequence of the inevitable coupling to the vacuum field
\cite{Scully:book}, it cannot be \textquotedblleft engineered
away\textquotedblright\ and must eventually be dealt with, in all QC
proposals. Several methods have been designed to this end, that may
roughly be classified as \textquotedblleft hardware\textquotedblright\
and \textquotedblleft software\textquotedblright :\ In the former
category are proposals to construct quantum computers in materials
where SE is strongly suppressed, e.g., placing atomic qubits in a
photonic band-gap structure \cite{Vats:99}. In the
latter category are various error correction, avoidance, and
suppression methods \cite{Mabuchi:96,Cirac:96,Plenio:97a,Alber:01,KhodjastehLidar:02,Beige:00,Agarwal:01}. With the exception of the $2\pi $ pulsing method of
\cite{Agarwal:01}, a\ unifying theme of these methods is to place the
system under continuous observation. It is then well known that the
Markovian quantum master equation can be unravelled into a set of
quantum trajectories, consisting of a conditional evolution (governed
by a non-Hermitian conditional Hamiltonian $H_{c}$, defined below),
randomly interrupted by quantum jumps (wavefunction collapse) into
different observed decay channels
\cite{Dalibard:92,Gardiner:book,Carmichael:93,Plenio:98}. The time
evolution conditional to a \emph{given} set of time-ordered
observations is called \textquotedblleft \textit{a posteriori}
dynamics\textquotedblright\ \cite {Barchielli:91}, and is \emph{not}
Markovian. The continuous observation can lead to a Zeno-effect type
suppression of decoherence, a fact that was exploited in
\cite{Beige:00}, in conjunction with an encoding into a
decoherence-free subspace (DFS) \cite{Zanardi:97c,Bacon:99a}, in order
to resist SE. Quantum error correcting codes (QECCs) can correct both
the conditional evolution and the jumps \cite{Cirac:96}, but more
efficient constructions are possible when one considers subspaces of
the full system's Hilbert spaces that are \emph{invariant under the
conditional evolution}.  It is then necessary to correct only the
errors arising due to the quantum jumps
\cite{Mabuchi:96,Cirac:96,Plenio:97a,Alber:01,KhodjastehLidar:02}. The
first proposal along these lines, \cite{Mabuchi:96}, did not consider
QC. A simple, but non fault-tolerant QC scheme, encoding a logical
qubits into two physical qubits (four atomic levels), tailored to SE
of phonons in trapped-ion QC, was subsequently presented in
\cite{Cirac:96}. A QECC correcting one arbitrary single-qubit error
and invariant under $H_{c}$ was given in \cite{Plenio:97a}, using an
encoding of one logical qubits into $8$ physical qubits. When one
makes the assumption that the qubit undergoing the quantum jump can be
identified (\textquotedblleft detected-jump\textquotedblright ), a
more efficient encoding is possible. A family of such ``detected-jump
codes'' (DJC)  was first
developed in \cite{Alber:01}, using a DFS to construct a subspace
invariant under $H_{c}$. In \cite{KhodjastehLidar:02} we showed how to
perform fault-tolerant universal QC on a subclass of such codes
encoding $n-1$ logical qubits into $2n$ physical qubits.

Here we present a new method for reducing and correcting SE errors.
Rather than constructing a code subspace invariant under $H_{c}$, we
dynamically eliminate $H_{c}$ by applying dynamical decoupling (or ``bang-bang'', BB) pulses
\cite{Viola:98,ShiokawaLidar:02}. We then construct a QECC that deals
with the remaining jump errors, under the detected-jump
assumption. The advantage of this method compared to the previous
methods using encoding is that it is significantly more economical in
qubit and pulse timing resources: It uses a QECC in which $n$ logical
qubits are encoded into only $n+1$ physical qubits; and, while\ in
\cite {Agarwal:01} the pulse interval has to satisfy the standard BB
condition of being shorter than the inverse of the bath high-frequency
cutoff \cite {Viola:98,ShiokawaLidar:02}, in our case the requirement
is that the pulses are faster than the average time between photon
emission events, which can be orders of magnitude longer. Furthermore,
our method is fully compatible with universal QC using Hamiltonians
that are naturally available in a large variety of QC proposals
\cite{LidarWu:02}, so unlike \cite{Vats:99} does not rely on one
specific architecture.

The idea of using a hybrid BB-encoding approach to suppress
decoherence was first proposed in \cite{ByrdLidar:01a}, where it was
pointed out that BB is fully compatible with encoding into a QECC or
DFS. In particular it was observed there that one could use BB to
suppress phase-flip errors, thus leaving the QECC\ with the need only
to correct bit-flip errors. However, no method specifically tailored
for SE errors was given. An experimental NMR\ implementation of a
hybrid BB-QECC was presented in \cite{Boulant:02}, where decoupling
was used to remove coherent scalar coupling between protons
(environment) and carbon qubits, together with QECC used to further
correct for fast relaxation due to dipolar interactions modulated by
random molecular motion.

Clearly, correcting for SE errors is only a part of a general
procedure for offsetting decoherence, as additional decoherence
sources will inevitably be present in any QC implementation. The
methods we present here therefore will have to become part of this
more general procedure, either as a first level of defence (in the
case that SE is dominant), or at higher levels in a concatenated QECC
scheme \cite{Gottesman:97a}, after other, more dominant errors have
been accounted for. The importance of the results presented here lies
in the fact that SE is \emph{always} present and therefore can never
be ignored. A code that is optimized with respect to this type of
error can potentially offer flexibility and significant savings in
resources and overhead.

The structure of this paper is the following. In Section~\ref{BB} we
show how the conditional evolution during SE can be
eliminated using a sequence of simple, global BB decoupling pulses. In
Section~\ref{QECC} we construct a simple and economical QECC that
corrects for the remaining quantum jump errors. We address fault
tolerance and various imperfections in Section~\ref{FT}. We then show
how to quantum compute in a universal and fault tolerant manner over
our QECC, using a variety of model Hamiltonians pertinent to a wide
class of promising quantum computing proposals. We conclude in Section~\ref{concl}.

\section{Eliminating the Conditional Evolution of Spontaneous Emission with
  BB pulses}
\label{BB}

Consider $N$ qubits that can each undergo SE, under the
 detected-jump assumption. This localizability of the SE events
 implies that the mean distance between
 qubits exceeds the wavelength of the emission.
 Note that this \emph{optical} distinguishability between qubits does not
 limit our ability to couple the qubits via non-optical interactions,
 of the type we consider in Section~\ref{FT-QC} below.

The ground and excited
 states of each qubit are denoted by $\left\vert 0\right\rangle $ and
 $\left\vert 1\right\rangle $ respectively. Let $\sigma
 _{i}^{-}=\left\vert 0\right\rangle _{i}\!\!\left\langle 1\right\vert
 $ denote the SE error generator acting on the $i$th
 qubit and let $\kappa _{i}$ denote the corresponding error rate. We
 use the quantum trajectories approach
 \cite{Dalibard:92,Gardiner:book,Carmichael:93,Plenio:98} to describe
 the dynamics of the decohering system. The evolution is decomposed
 into two parts: a conditional \emph{non-Hermitian} Hamiltonian
 $H_{c}$, interrupted at random times by occurrence of random jumps,
 each corresponding to an observation of decay channels in a quantum
 optical setting. For errors such as SE, where the jump can be
 detected by observation of the emission, the quantum trajectories
 approach also provides us with a way to combine QECCs and BB, in
 analogy to the way this was done for QECC and DFS in \cite{Alber:01,KhodjastehLidar:02}. The BB pulses take care of the
 conditional evolution, whereas the QECC deals with the random
 jumps. The conditional Hamiltonian is given in the SE case by \cite {Dalibard:92,Gardiner:book,Carmichael:93,Plenio:98}:
 $H_{c}=-\frac{i}{2} \sum_{i=1}^{N}\kappa _{i}\sigma _{i}^{+}\sigma
 _{i}^{-}$, where $\sigma _{i}^{+}=\left( \sigma _{i}^{-}\right)
 ^{\dag }$. In \cite{KhodjastehLidar:02} we assumed that the
 environment effectively does not distinguish among the qubits that
 undergo SEs ($\kappa _{i}=\kappa $) and the
 conditional Hamiltonian would then become: $-\frac{i}{2}\kappa
 \sum_{i}\left\vert 1\right\rangle _{i}\!\!\left\langle 1\right\vert
 $. This assumption is \emph{not} necessary in the current
 work. From here on operators $X_{i},Y_{i},Z_{i}$ refer to the
 corresponding Pauli matrices acting on the $i$th qubit, and $I$
 denotes the identity matrix. Now suppose that we apply a collective
 $X\equiv \otimes _{j=1}^{N}X_{j}$ pulse to the system, at
 intervals $T_{c}/2\ll $ $ 1/\gamma $, where $\gamma $ is
 the SE rate.\footnote{We re-emphasize that this time can be much
 longer than the bath correlation time typically assumed to set the
 time-scale for BB operations (for an exception see
 \cite{ShiokawaLidar:02}).} Under this condition, and using
 $X_{i}\sigma _{i}^{-}X_{i}=\sigma _{i}^{+}$ we can write the
 evolution after a full $T_{c}$ period as:
\begin{eqnarray*}
U & =&\exp (-i\frac{T_{c}}{2}H_{c})X\exp (-i\frac{T_{c}}{2}H_{c})X \\
& =& \exp (-\frac{T_{c}}{4} \sum_{i}\kappa_i\left\vert 1\right\rangle
\!\!_{i}\left\langle 1\right\vert )\exp (-\frac{T_{c}}{4}
\sum_{i}\kappa_i\left\vert 0\right\rangle \!\!_{i}\left\langle
0\right\vert ) \\ & =& \exp (-\frac{T_{c}}{4}\sum_i\kappa_i )I,
\end{eqnarray*}
where $I$ is the identity operator. Therefore the decohering effect of
the conditional Hamiltonian (that distinguishes states with different
numbers of $1$'s) is removed and replaced by an \emph{overall}
shrinking norm. When the jumps are included in the dynamics, the state
must be renormalized
\cite{Dalibard:92,Gardiner:book,Carmichael:93,Plenio:98}, so this
shrinking disappears. Note that we have \emph{not} eliminated
Markovian decoherence using BB pulses, since we have considered only
a single trajectory. In fact, a comparison of the coherence
$\mathcal{C}=\mathrm{Tr}{\left( \rho ^{2}\right) }$ (where $\rho $ is
the qubit density matrix) shows that if the results are
ensemble-averaged over the \textit{a posteriori} dynamics (recovering
the Markovian master equation), and the jump errors are not corrected,
then there is no advantage in using a BB sequence. More specifically,
when comparing $\mathcal{C}$ for the (1)\ free and (2)$\ $ every
$T_{c}/2\ X$-pulsed evolution of a single qubit undergoing SE with
rate $ \gamma $, we find:
\begin{eqnarray*}
\mathcal{C}_{1}& = &1-\gamma T_{c}(\beta ^{2})+O(\gamma ^{2}) \\
\mathcal{C}_{2}& = &1-\gamma T_{c}(\alpha ^{4}+\beta ^{4})+O(\gamma
^{2})
\end{eqnarray*}
where the initial qubit state $\left\vert \psi \right\rangle =\alpha
\left\vert 0\right\rangle +\beta \left\vert 1\right\rangle $ is
normalized: $ \alpha ^{2}+\beta ^{2}=1$. Averaging over a random
sample of initial states chosen from a uniform distribution (with
$\alpha$ and $\beta$ subject to normalization), we have
$\exv{\mathcal{C}_{1}}=\exv{\mathcal{C}_{2}}$ , so as expected for
purely Markovian dynamics, there is no improvement after using
\emph{just} BB pulses.

\section{Correcting Spontaneous Emission Jumps with a QECC}
\label{QECC} We now introduce a very simple QECC that corrects the remaining part of the
decoherence process, the random jumps.  Since the error correction process by necessity takes place
during the conditional evolution (the jump is instantaneous and the QECC\ takes time), we must
ensure that the QECC keeps its error correcting properties under the conditional Hamiltonian and BB
pulses. A minimal example of such a \textquotedblleft decoupled-detected jump
corrected\textquotedblright\ code is given by a subspaces of the $ N=n+1$ qubit Hilbert space
$\mathcal{C}_{n}$,  spanned by the codewords
\begin{align}
|x\rangle _{L}&\equiv |{x_{1},\ldots ,x_{n}}\rangle
 _{L}\nonumber\\&=\frac{|{x_{1},\ldots ,x_{n}},0\rangle+ |\overline{x_{1}},\ldots ,\overline{x_{n}},1\rangle}{\sqrt{2}}.
\label{eq:code}
\end{align}
where $\{x_i\}$ is the binary representation of the $n$-qubit state $\ket{x}$ and $\overline{x}$ is
an inverted $x$ in which $1$ and $0$ are interchanged. For example, for $n=2$, the code
$\mathcal{C}_{2}$ is (up to normalization by factors of $\sqrt{2}$):
\begin{eqnarray}
\left\vert \underline{00}\right\rangle _{L} &=&\left\vert {00}0\right\rangle+ \left\vert
{11}1\right\rangle \quad \left\vert \underline{01}\right\rangle _{L}=\left\vert {01}0\right\rangle+
\left\vert {10}1\right\rangle \notag \\ \left\vert \underline{10}\right\rangle _{L} &=&\left\vert
{10}0\right\rangle+ \left\vert {01}1\right\rangle \quad \left\vert \underline{11}\right\rangle
_{L}=\left\vert {11}0\right\rangle+ \left\vert {00}1\right\rangle \notag  .  \label{eq:C_2}
\end{eqnarray}
That $\mathcal{C}_{n}$ is a QECC against the jump errors follows from the fact that a spontaneous
emission error at a given qubit position $i$ eliminates the component of the codeword with $0$ in
that position, which by construction results in a surjective mapping between the original codewords
and the resulting states that are orthogonal to each other. More specifically, the sufficient
condition that a QECC must satisfy is that orthogonal codewords must be mapped to orthogonal states
after the occurrence of errors, so that the errors can be resolved and undone \cite {Knill:97b}.
Recall that here we are assuming that we know the location of the error, after recording the
position of the SE. Hence we need only compare orthogonal codewords after the
action of an error in a \emph{known} location $i$:
\begin{equation*}
_{L}\langle y|\sigma _{i}^{+}\sigma _{i}^{-}|x\rangle _{L}=\left\{
\begin{array}{c}
\delta _{xy}/2\quad \mathrm{if}\text{ }y_{i}=x_{i} \\ 0\quad
\mathrm{if}\text{ }y_{i}\neq x_{j}
\end{array}
\right.
\end{equation*}
The second line is explained in the following way: If $x_i\neq y_i$, then either $x_i$ or $y_i$ is
$0$. Suppose $y_i$ is zero, then the component of the $\ket{y}_L$ codeword that remains after the
SE is $\sigma _{i}^{-}\ket{\overline{y},1}$ and the component of $\ket{x}_L$ that remains is
$\sigma _{i}^{-}\ket{{x},0}$ which are always orthogonal to each other. Thus the QECC condition is
satisfied. To see that recovery from the errors is indeed possible, we describe a simple (non
fault-tolerant) scheme. The recovery operation after the detection of an error in position $i$, is
given by $U=\prod_{n\neq i}X_{n}\prod_{n\neq i}\textsc{CNOT}_{in}H_{i}$, where $H_{i}$ is a
Hadamard operation on qubit $i$ and $\textsc{CNOT}_{ij}$ is a \textsc{CNOT} gate with qubit $i$
($j$) as the control (target), i.e., $\textsc{CNOT}_{ij}|x_{i},x_{j}\rangle =|x_{i},x_{j}\oplus
x_{i}\rangle $. That this unitary operation fixes the SE error can be seen as follows by
considering the transformation of the codewords after the error and recovery for the two cases:
$x_i=1$ and $x_i=0$. If $x_i=1$ ($x_i=0$) then the codeword after the error becomes $\ket{x_1\ldots
0 \ldots x_{n+1}}$ ($\ket{\overline{x_1}\ldots 0 \ldots \overline{x_{n+1}}}$). It is easy to verify
that applying $U$ to this state indeed returns the original logical codeword $\ket{x}_L$. To
illustrate this we discuss in detail the case of $\mathcal{C}_{2}$. The conditional evolution,
under the collective BB pulse $X=X_{1}X_{2}X_{3}$, has the sole effect of shrinking the norm of all
codewords in Eq.~(\ref{eq:C_2}) \emph{equally}. Thus the BB-modified conditional evolution does not
change the orthogonality of the codewords. Now suppose SE from the first qubit has been observed.
Then an arbitrary encoded state $|\psi \rangle _{L}=a\left\vert 00\right\rangle _{L}+b\left\vert
01\right\rangle _{L}+c\left\vert 10\right\rangle _{L}+d\left\vert 11\right\rangle _{L}$ changes
into $\ket{\psi_{\text{err}}}=a\ket{011}+b\ket{001}+c\ket{000}+d\ket{010}$. To reverse the error we
use the unitary operator $U=X_2X_3\textsc{CNOT}_{12}\textsc{CNOT}_{13}H_1$. The erred state is then
transformed to $ U|\psi _{\mathrm{err}}\rangle
=(1/\sqrt{2})(a\ket{00}_L+b\ket{01}_L+c\ket{10}_L+d\ket{11}_L)=\ket{\psi}_L$.

\section{Fault Tolerant Preparation, Measurement, and Recovery}
\label{FT}
So far we have assumed perfect error detection, recovery, and
gates. Of course, in reality these assumptions must be relaxed. Here
we discuss the implications of imperfections.

In general, a procedure is said to be fault tolerant if the occurrence
of an error in one location does not lead (via the applied procedure)
to the catastrophic multiplication of errors in other
locations \cite{Gottesman:97a}, an event that the code cannot
correct.

Let us first discuss \emph{preparation} of the encoded qubits. The code word $|0\rangle_{L}
$ is prepared by cooling all physical qubits in their ground
state ($\ket{0}$), which can be done, e.g., via cooling, a strong polarizing field, or repeated
strong measurements of all qubits, followed by
a Hadamard on the $n+1$-th qubit, and
a collective CNOT from the $n+1$-th qubit to all the
other physical qubits $1$ to $n$. Once $|0\rangle _{L}$ has been prepared computation proceeds
using the fault-tolerant logical operations given in Section~\ref{FT-QC} below, so any other state
can be reached fault-tolerantly. \emph{Readout} is also simple: First apply the same collective
\textsc{CNOT} and then measure the first $n$ physical qubits. The measurement procedure must be
tailored to the specific implementation, but our only assumption is that single-qubit measurements
are possible, and that these measurements do not couple qubits. The measurement procedure is then
fault-tolerant. If means of applying fault-tolerant \textsc{CNOT} are available then both
preparation and readout are fault tolerant.

Next consider \emph{recovery}. The code $\mathcal{C}_{n}$ is an especially simple example of CSS
stabilizer codes \cite{Gottesman:97}, with stabilizer generated by the single element $\otimes
_{j=1}^{n+1}X_{j}$. It is well known how in general to perform fault-tolerant recovery from this
class of codes \cite{Gottesman:97a} (see also \cite{Lidar:00b}), so we will not repeat the general
construction here, which involves preparing and measuring encoded ancilla qubits (note that this
typically doubles the number of physical qubits required, even before concatenation).

Finally consider \emph{detection of SE events}. Above we assumed that it is possible to perfectly identify the position of
a qubit that underwent SE. Note that this measurement is in itself
fault-tolerant, in the sense that observing an SE event on a specific
qubit cannot cause errors to multiply. Clearly, detecting which qubit emitted a photon is very demanding
experimentally, and can in practice only be done to some finite
precision (though there is no fundamental limit, provided the distance
between the qubits is larger than the wavelengths of emitted photons),
and at the cost of introducing potentially cumbersome detection
apparati. The same difficulty is shared by previous ``detected-jump''
schemes \cite{Alber:01,KhodjastehLidar:02,Beige:00}.

More specifically, in reality there is a
finite probability that the emitted photon will (i) Go undetected;
(ii) Be attributed to the wrong atom (misidentification). The latter
possibility applies also to other qubit measurements; (iii) In case
(ii), there is the additional possibility of an error by applying the correction step to the
wrong qubit. In general, fault-tolerance results again come to the
rescue: provided that the probability of an undetected photon and/or
misidentification can be kept sufficiently low, concatenated QECC
guarantees that the procedure will remain robust
\cite{Gottesman:97a,Knill:98,Steane:02}. However, several additional
comments are in order. First, we note that the performance of DJC codes in the presence
 of imperfections such as detection inefficiencies and time delay between error detection and
 recovery operations, has been analyzed in \cite{Alber:02a}, with
 favorable conclusions regarding fidelity degradation. We expect
 similar conclusions for our current  method. Second, unlike the case
 of DJC codes \cite{Alber:01,KhodjastehLidar:02,Alber:02a}, we do
 not require equal error rates $\kappa _{i}$. Hence our qubits need
 not be identical: qubits can be tuned to different cavity modes and
 therefore emit distinguishable photons. This should enable a
 significant reduction in the misidentification error rate. Third, we
 can take advantage of the fact that after any SE event, each codewords is transformed
 to a state which is orthogonal to all original codewords. Thus, we can perform an extra measurement
 (of the stabilizer $X_1\ldots X_{n+1}$) that determines whether an error has occurred at all.
This is done by adding one more ancilla qubit $a$ (initalized into
 the $|0\rangle$ state) that functions as a syndrome-measurement bit. Now repeatedly apply $\prod_{i=1}^{n+1} H_i \text{CNOT}^i_a H_i$
and periodically check the qubit $a$ to see if it has changed to
 $|1\rangle$. In such a case if the position of the SE is undetected the computation has to be
 restarted and the ancilla qubit has to be reset; otherwise the recovery procedure may still be applied.
Fault tolerant procedures are known for syndrome measurement as
 well \cite{Gottesman:97a,Knill:98,Steane:02}. There is also the possibility
of SE on the ancilla qubit, but this can only be caused by two successive spontaneous emissions
(one on the code qubits and one on the parity qubit), which has a lower probability $p^2$, where
$p$ is the probability of two SE errors occurring during the same observation period, before the
first one is detected. Note that the parity bit also helps preventing the error of applying a
correction step without an SE event having taken place.

\section{Fault Tolerant Computation}
\label{FT-QC}
So far we have described a fault tolerant implementation of
quantum \emph{memory} in the presence of SE. Now we describe how to
perform universal quantum \emph{computation} fault tolerantly on our
code. Formally, one can use the formalism of normalizer group
operations, together with a non-normalizer element such as the $\pi
/8$ or Toffoli gate \cite {Gottesman:97a}. However, here we are
interested in how to carry this out from the perspective of \emph{the
  naturally available interactions for a given
  physical system}. Similar
questions have been raised recently under the heading of
\textquotedblleft encoded universality\textquotedblright :\ the
ability to quantum compute universally directly in terms of a given
and limited set of Hamiltonians, possibly by use of encoded qubits
(see, e.g., \cite{LidarWu:01,Kempe:01} and references therein). The
problem then translates into finding sets of Hamiltonians that
generate a universal set of logic gates on the code. There are many
options, depending on the set of naturally available
interactions. Nevertheless, all encoded universality constructions
rely on showing that the well-known universal set of all single-qubit
operations and a single entangling gate can be generated, on the
encoded qubits.  Underlying this are a few elementary identities. Let
us define \emph{conjugation} as:
\begin{equation*}
A\overset{B,\varphi }{\rightarrow }\equiv e^{-i\varphi B}Ae^{i\varphi
B}.
\end{equation*}
Then for any three $su(2)$ generators $\{J_{x},J_{y},J_{z}\}$:
\begin{equation}
J_{x}\overset{J_{z},\varphi }{\longrightarrow }J_{x}\cos \varphi
+J_{y}\sin \varphi .  \label{eq:su2}
\end{equation}
This can be lifted to unitary evolutions using
\begin{equation}
Ue^{A}U^{\dagger }=e^{UAU^{\dagger }}, \label{eq:lift}
\end{equation}
valid for any unitary $U$. Hence where convenient we present our
arguments in terms of transformed Hamiltonians.
Eqs.~(\ref{eq:su2}),(\ref{eq:lift}) show that given two $su(2)$
generators one can generate a unitary evolution about any axis. This
is also the basis for the well-known Euler angle construction, used to
argue that all single qubit operations can be generated from $\sigma
^{x}$ and $\sigma ^{z}$ Hamiltonians:\ an arbitrary rotation by an
angle $\omega $ around the unit vector $\mathbf{n}$ is given by three
successive rotations around the $z$ and $x$ axes: $e^{-i\omega
\mathbf{n}\cdot \mathbf{\sigma }}=e^{-i\beta \sigma ^{z}}e^{-i\theta
\sigma ^{x}}e^{-i\alpha \sigma ^{z}}$
\cite{Nielsen:book}. Eqs.~(\ref{eq:su2}),(\ref {eq:lift}) show that
this is true also for \textquotedblleft encoded
Hamiltonians\textquotedblright , which we define as Hamiltonians that
have the same effect on encoded states as do regular Hamiltonians on
\textquotedblleft bare\textquotedblright\ (unencoded) qubits. We
denote encoded Hamiltonians by a bar. For the code states
(\ref{eq:code}) these are given by:
\begin{equation}
\bar{Z}_{i}=Z_{i}Z_{n+1},\quad \bar{X}_{i}=X_{i},
\end{equation}
and generate $su(2)$. Therefore controllable $Z_{i}Z_{n+1}$ and $X_{i}$ Hamiltonians suffice to
generate arbitrary single encoded-qubit transformations. To complete the set of universal logic
gates we require \emph{some} non-trivial (entangling) gate \cite{Dodd:01}, such as
controlled-phase:\ $CP=\mathrm{diag}(1,1,1,-1)$, in the computational basis.  $CP$ can be generated
from the Ising interaction $Z_{i}Z_{j}$ as follows:\ $ CP_{ij}=e^{-i\frac{\pi
}{4}(Z_{i}+Z_{j})}e^{-i\frac{3\pi }{4}Z_{i}Z_{j}}$. An entangling gate can also be generated from
the Hamiltonian $X_{i}X_{j}\,\ $[one way to see this is to note that it can be rotated to
$Z_{i}Z_{j}$ using $Y_{i}$ and $Y_{j}$ in Eqs.~(\ref{eq:su2}),(\ref{eq:lift})]. Encoded $ CP$ can
thus be generated from the encoded Hamiltonians $\bar{Z}_{i}\bar{Z} _{j}=Z_{i}Z_{j}$ or
$\bar{X}_{i}\bar{X}_{j}=X_{i}X_{j}$. Note that in both cases the physical interaction is also the
corresponding encoded Hamiltonian. Thus the sets of controllable Hamiltonians $\{X_{i},Z_{i}Z_{j}\}
$ or $\{X_{i},Z_{i}Z_{n+1},X_{i}X_{j}\}$ suffice for encoded universal QC on our code. Importantly,
these sets moreover exhibit \textquotedblleft natural fault-tolerance\textquotedblright\
\cite{Bacon:99a}: they \emph{preserve the code subspace} and hence will not expose the code to
uncorrectable errors. An \emph{accurracy} error in the time over which the Hamiltonians are turned
on can be dealt with using the technique of concatenated QECCs \cite {Gottesman:97a}. The question
now is how to generate these sets, or an equivalent fault tolerant universal set, from the given,
naturally available interactions. We will consider here the most important cases, extending methods
developed in \cite{LidarWu:01,WuLidar:01,WuLidar:01a}. Note that the decoupling procedure requires
us to assume in any case the ability to apply a \emph{global} (non-selective) $X$ pulse, and the
recovery procedure requires the ability to apply a \textsc{CNOT} gate. We comment on these
requirements in each of the cases we next analyze.

\subsection{Case 1: Natural $\{Z_{i}$, $X_i, X_{i}X_{j}\}$}
The Hamiltonians $Z_{i},X_{i}X_{j}$ are naturally available, e.g., in the S{\o }rensen-M{\o }lmer
scheme for trapped-ion QC \cite{Sorensen:00}, and in proposals using Josephson charge qubits
\cite{You:02}. However these do not form a universal set for our code and hence we must assume the
ability to turn on spin-selective $ X_{i}$ Hamiltonians. This will also be sufficient for producing
the encoded $Z_iZ_j$ coupling.

\subsection{Case 2: $\{Z_{i}, X_i, XY$ Model$\}$}
Members of a relatively large class of promising QC\ proposals (quantum dots\ \cite
{Imamoglu:99,Quiroga:99}, atoms in a cavity \cite{Zheng:00}, quantum Hall qubits
\cite{Mozyrsky:01}, subradiant dimers in a solid host \cite {Petrosyan:02}, capacitively coupled
superconducting qubits \cite{Vion:02})\ have a controllable Hamiltonian of the XY form: $
H_{ij}^{XY}=J_{ij}(X_{i}X_{j}+Y_{i}Y_{j})$. Let $T_{ij}\equiv \frac{1}{2} (X_{i}X_{j}+Y_{i}Y_{j})$.
Then $|01\rangle \overset{T_{12}}{ \longleftrightarrow }|10\rangle $, and annihilates $|00\rangle
,|11\rangle $. I.e., the XY Hamiltonian cannot change the total number of $1$'s in a computational
basis state \cite{WuLidar:01,WuLidar:01a}. Therefore by itself, or even if supplemented with
$Z_{i}$ Hamiltonians, it cannot generate $su(2)$ on our code. This conclusion is unchanged even if
one considers conjugating $H_{ij}^{XY}$ with $H_{ik}^{XY}$: then $
\{T_{12},T_{13},-Z_{1}Z_{2}T_{23}\}$ close as $su(2)\,$, and still preserve the total number of
$1$'s. Therefore in this case we must assume the ability to tune $X_{i}$ Hamiltonians as well, to
obtain universality.
Now, $X_{i}X_{j}(T_{jk})X_{i}X_{j}=\frac{1}{2} (X_{j}X_{k}-Y_{j}Y_{k})$, which commutes with
$T_{jk}$. Therefore, using Eq.~(\ref{eq:lift}), we have $X_{i}X_{j}e^{-i\theta
T_{jk}}X_{i}X_{j}e^{-i\theta T_{jk}}=e^{-i\theta X_{j}X_{k}}$, showing that the \emph{Hamiltonian}
$X_{j}X_{k}$ can be generated in four steps. At this point we have the same set of Hamiltonians as
in Case 1, so that universal encoded computation is possible, as are the global $X$ pulse and
recovery.

\subsection{Case 3: $\{Z_{i}, X_i$, Heisenberg interaction$\}$}

Next we consider the case of single-qubit $X$-$Z$ control together with the Heisenberg interaction
$H_{ij}^{\mathrm{Heis}}=J_{ij}(X_{i}X_{j}+Y_{i}Y_{j}+Z_{i}Z_{j})$.  Heisenberg interactions prevail
in QC\ proposals using spin-coupled quantum dots \cite {Loss:98,Levy:01a,Friesen:02} and donor
atoms in Si \cite{Kane:98,Vrijen:00} . This case is similar to that of the XY\ model, since
$H_{ij}^{\mathrm{Heis} }$ also preserves the total number of $1$'s in a computational basis state.
Therefore, as in the XY\ case, we must assume the ability to generate an $ X_{i}X_{j}$
\emph{pulse}. Then, $X_{i}X_{j}(H_{jk}^{\mathrm{Heis}
})X_{i}X_{j}=J_{jk}(X_{j}X_{k}-Y_{j}Y_{k}-Z_{j}Z_{k})$, which commutes with $
H_{jk}^{\mathrm{Heis}}$, so that $X_{i}X_{j}e^{-itH_{jk}^{\mathrm{Heis}
}}X_{i}X_{j}e^{-itH_{jk}^{\mathrm{Heis}}}=e^{-2itJ_{jk}X_{j}X_{k}}$, and we are back to Case 1.
There is now another option for generating an entangling gate: we can generate a pure $ZZ$
interaction using $ZI(H_{\mathrm{Heis} })ZI=-XX-YY+ZZ$, which commutes with $H_{\mathrm{Heis}}$, so
that $e^{-itH_{ \mathrm{Heis}}}e^{-i\frac{\pi }{2}ZI}e^{-itH_{\mathrm{Heis}}}e^{-i\frac{\pi
}{2}ZI}=e^{-2itJZZ}$. This is a four-step, naturally fault tolerant procedure. The decoupling pulse
and recovery are now the same as in Case 1.

Finally, there remains the issue of compatibility between the encoded logic operations and the
decoupling pulses that are being constantly applied to the system. All three interaction
Hamiltonians we have considered commute with the global $X$ BB-pulse, so are fully compatible with
the BB operations. Furthermore, the logical single-qubit terms also commute with the $X$ pulse. Thus
whenever use of a single body $Z_i$ Hamiltonian is required, it must be synchronized to be
applied only after an even number of collective BB pulses, to ensure the compatibility of quantum
manipulation and dynamical decoupling.

\section{Conclusions}
\label{concl}
We have proposed a new method for performing universal, fault tolerant
quantum computation in the presence of spontaneous emission. The
method combines dynamic decoupling pulses with a particularly simple
and efficient quantum error correcting code, encoding $n$ logical
qubits into $n+1$ physical qubits. Computation is performed by
controlling single-qubit $ \sigma ^{x}$ and $\sigma ^{z}$ terms
together with any of three major examples of qubit-qubit interaction
Hamiltonians, applicable to a wide range of quantum computing
proposals. The proposed method offers an improvement over previous
schemes for protecting quantum information against spontaneous
emission in that the code is at least twice as efficient in terms of
qubit resources, and the method is fully compatible with computation
using physically reasonable resources and interactions.

\begin{acknowledgments}
This material is based on research sponsored by the Defense Advanced
Research Projects Agency under the QuIST program and managed by the Air
Force Research Laboratory (AFOSR), under agreement F49620-01-1-0468. D.A.L. further gratefully acknowledges
financial support from Photonics Research Ontario. We thank
Dr. D. Bacon and Dr. S. Schneider for useful discussions.
\end{acknowledgments}


\end{document}